# AUTONOMY AT LEVELS FOR SPACECRAFT

Daniel Baker*, Jeremy Wojcik†, and Sean Phillips‡§¶

Autonomy at Levels is the idea that autonomy should be embedded within and throughout a spacecraft. Using Systems Engineering methods a spacecraft is typically decomposed into systems, subsystems, assemblies, components, and so on. All these decomposition levels within all the spacecraft's systems, could and should have autonomy elements built in. As a result, the "autonomy system" is made of autonomy elements or units that are integrated, distributed and embedded within the whole spacecraft. This is like how the power system would be designed and implemented. Linking control loops and autonomy loops illustrates how to achieve Autonomy at Levels.

## INTRODUCTION

Partially inspired by the human nervous system and partially by our background in control systems, we see autonomy as a designed-in element of all machines, and autonomy elements at all levels of the system decomposition. In previous work ([4], [3]), we defined Levels of Autonomy for spacecraft, clusters, constellations, the ground segment, and the user segment. These are top-down definitions. We now turn our attention to a bottom-up approach of developing Autonomy at Levels.

The biological nervous systems, in the animal kingdom, have many levels of autonomous functionality. At lower-levels, axon reflexes happen locally without higher-levels, like the spinal cord and brain, being involved. At mid-levels, reflex arcs (loops) involving the spinal cord and motor neuron with somatosensory feedback coordinates most autonomic nervous system (ANS) functions. In normal operation, these loops are handled without direct brain involvement, but the higher-level inputs can influence, inhibit, or excite lower-level normal functioning, as needed. At the lowest levels, inside a biological neuron, a control system is active trying to keep the cell of the neuron at homeostatic conditions. In traditional control theory, this homeostasis process is called a regulator. Many control loops work at the lowest levels within a spacecraft system and are mimicking biology, either intentionally or unintentionally. We embrace this inspiration from biology in our autonomy model for spacecraft.

---

*Aerospace GNC Engineer, Defense, AV, Albuquerque, NM, USA.

†Mathematician/Physicist, Defense, AV, Albuquerque, NM, USA.

‡Technical Advisor, Space Control Branch (RVSW), Air Force Research Laboratory, Albuquerque, NM, USA.

**Biological Inspiration**

From Nestler in [18]:

> The overall function of the ANS is to maintain homeostasis in the body (i.e. optimize conditions for survival) in the face of constantly changing environmental and activity demands. For example, the ANS adjusts blood pressure and heart rate to meet the circulatory needs of the body that can vary tremendously from supine sleep to vigorous exercise. The ANS also maintains a constant body temperature despite changing environmental conditions and metabolic activity. Under ordinary circumstances, the ANS functions independently of consciousness yet can be influenced to some degree by volition and emotion.

The ANS is divided into three anatomically distinct divisions: sympathetic, parasympathetic, and enteric. Under normal conditions, the sympathetic and parasympathetic systems both simultaneously influence the the local organs, glands and vascular systems they manage. The level of control of each of these systems is based on the relative signal strength each is directing at the organ/gland/system and can range from low to high. High indicates dominant control while low indicates subordinate control. For example, when entering a dark room, the sympathetic neurons begin firing at a high rate toward the iris of the eye, causing it to dilate the pupil so that more light can enter while the firing of the parasympathetic nerve drops to very low levels. In this case sympathetic control is high/dominate and parasympathetic control is low/subordinate. When returning to brightly lit area parasympathetic neurons begin firing at a high rate causing the iris to constrict the pupil so that less light enters and firing of the sympathetic system decreases. (Adapted from [10].)

As explained in [21], the ANS uses a hierarchy of reflexes to control the function of autonomic target organs. These reflexes range from local, involving only a part of one neuron, to regional, requiring mediation by the spinal cord and associated autonomic ganglia, to the most complex, requiring action by the brain stem and cerebral centers. Nichols ([19]) and Veterian Key ([9]) also show three hierarchical levels of reflexes although they label the levels differently (Figure 1).

The local reflexes are refereed to as axon reflexes and are responsible for dilation of skin blood cells resulting in a localized red flare, stimulation of sudomotor nerve terminals that leads to sweating, and stimulation of adrenergic small fibers in the skin that causes goosebumps. The reflex arc of axon reflex has neither an integration center nor any synapse which are common in all other reflexes ([22]).

The regional reflexes do involve the spinal cord but do not need to go up to the brain for normal functioning. These reflexes may involve one, two, or a few synapses and are short-latency reflexes. Longer latencies commonly involve pathways that ascend to the brain stem or higher in the brain ([15]). These reflexes are responsible for two main functions in the body. One is to protect muscle from injury in case of instant involuntary movements. The other is to control fundamental rhythmic and repetitive movements like walking and running ([8]).

The simplest form of homeostatic sensory-control loop is the classical (regional) reflex arc. In reflexes, deviations between sensory inputs and internal setpoints trigger predefined hard-wired reactions. A classic example is the baroreflex. Here, a detected increase in blood pressure, signaled via baroreceptors, triggers a reaction in the spinal cord that results in a short-term down-regulation of blood pressure via barosensitive autonomic efferents. The simple reflex arc control loops are adaptive but limited. When the environment is dynamic and the default reaction is no longer beneficial the control mechanisms have to make reflexes more flexible by allowing temporarily movement away from its setpoint or by changing the setpoints themselves ([5]). Other setpoint examples include core body temperature and systemic sugar levels. These are regulated by reflex arcs without the involvement of higher brain areas. These types of classical reflex arcs are reactive systems that provide direct negative feedback control by triggering actions when parameters deviate from their predefined setpoints. Examples of more complex reflex arcs that allow for dynamic changes in setpoints include when exercise dampens the baroreceptor reflex to allow heart rate and blood pressure to remain elevated as needed ([14]).

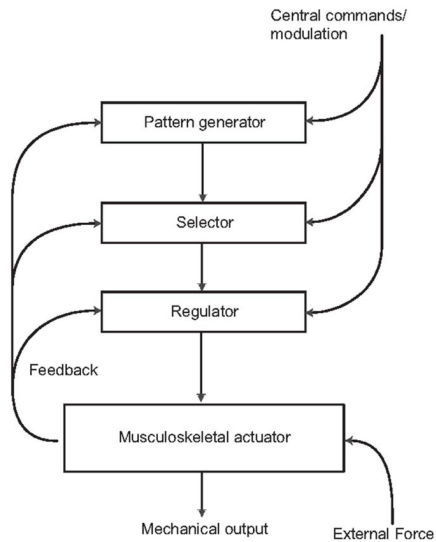 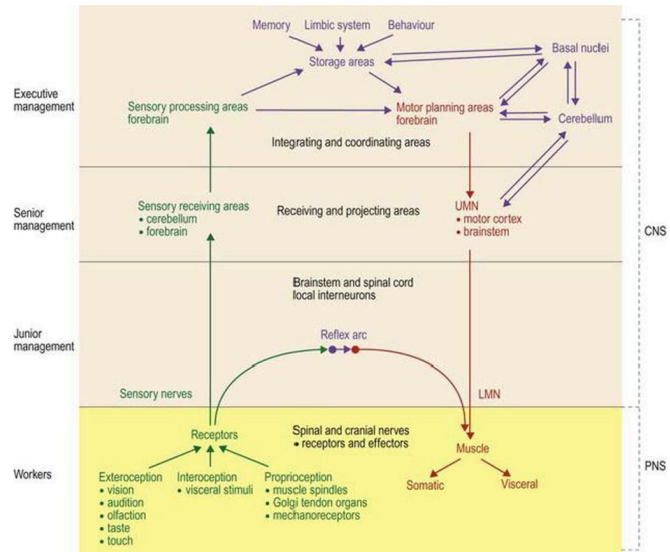

**Figure 1. Hierarchical Organization of Reflex Circuits ([19], [9])**

Central pattern generators (CPGs) are relatively small, relatively autonomous groups of neurons that produce the patterned, rhythmic outputs that regulate rhythmic and repetitive movements. In addition to walking and running, CPGs are also responsible for dancing, chewing, and swallowing. CPGs show that nervous systems can create outputs without sensory input and function in a manner similar to electronic oscillators in clocks and timing devices ([12]).

Reflexive activities can also involve more complex pathways, including additional neuronal relays. This additional processing of information within the brain (or brain stem) can lead to more nuanced responses based on other sensory stimuli or the physiological state of the tissue or organism

This inspiration from biology, indicates that machine autonomy need not have a single master brain that monitors and controls all aspects of the full system. In fact, following evolution's lead, a single master brain is not optimal. Ultimately, neuroscience, has not yet, but should inspire a more complete, multi-layered framework for autonomy and leads to embedded autonomy elements at all levels within the system decomposition. In Figure 2, the overall process for system decomposition, requirements flow down, build, integration, test, and verification and validation as generally used in Systems Engineering is shown.

Autonomy at Levels is the idea that autonomy should be an embedded within and throughout a spacecraft. Using Systems Engineering methods ([7], [11], [6]), a spacecraft is typically decom-posed into systems, subsystems, assemblies, components, and so on. This is sometimes reflected in the work breakdown structure (WBS) or product breakdown structure (PBS) ([2], [13]). All these decomposition levels within all the spacecraft's systems, could and should have autonomy elements built in. As a result, the "autonomy system" is made of many autonomy elements or units that are integrated, distributed and embedded within the whole spacecraft. This is like how the "power system" would be designed and implemented. The power system has connections and interfaces with just about every other piece, component, and subsystem and the autonomy system should as well.

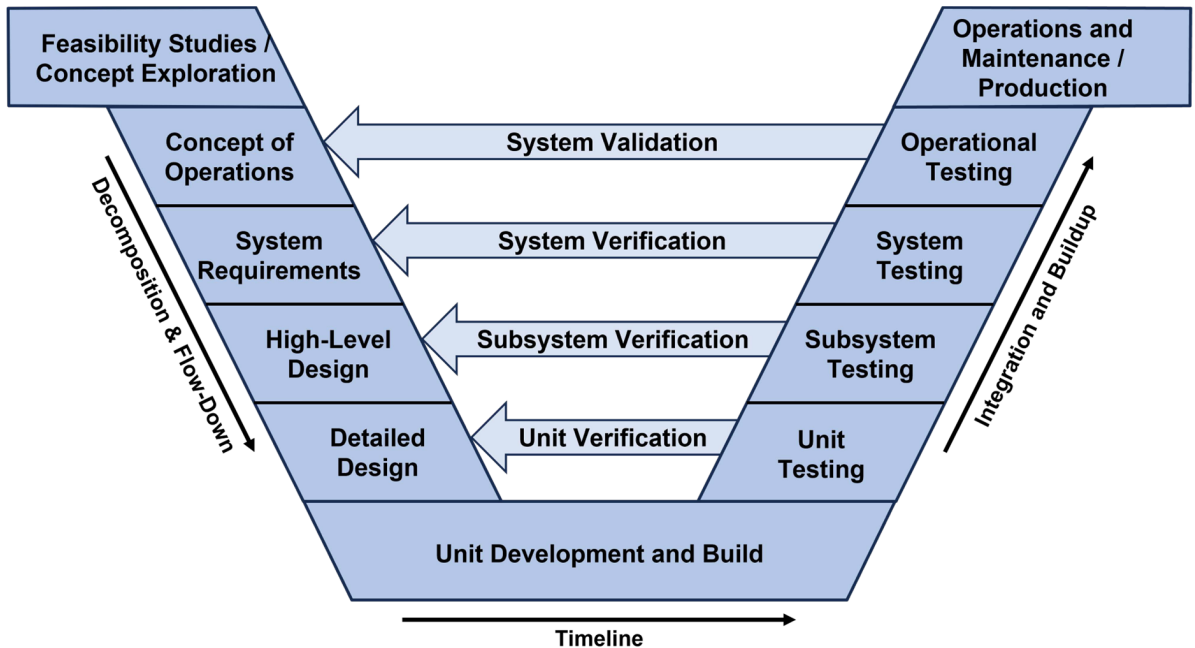

**Figure 2. Systems Engineering V**

**Outline**

In the following sections of this paper, a little background on machine autonomy, some definitions, and a review of autonomy's primary functions, relationships and interactions will be provided. Following that, more details about the idea of autonomy as an outer loop around each and every control loop within a spacecraft will be discussed. A spacecraft system breakdown taxonomy is introduced and is used in several simple examples to illustrate our ideas. The paper ends with a summary and some remaining questions and future work suggestions.

**BACKGROUND**

Autonomy means many things to many people but for most autonomy (for humans or machines) invokes the notion of independence. The ability to direct one's own decisions from personal life to interactions with the world. In other words, to control what one does, when one does it, and who one does it with. For machines, in particular, people primarily think of the system being able to function without (or with little) need for human intervention but for the purpose of doing some work for humans.

From basic physics, machines need to move to do work. The motion of objects is governed by physics and described by differential equations. Controls is the engineering discipline for regulating motion and autonomy should be the engineering discipline for regulating controls. That is, autonomy uses controls to execute actions to achieve its (autonomy's) goals and tasks. Autonomy is the who, what, when, and where to be done (the tasks) to achieve goals and controls is how it is done (the actions). Autonomy determines what needs doing (planning) and controls does the doing (execution). Autonomy functions, then, are naturally tied to control loops.

Autonomy is a continuous sliding scale from pure automation of repetitive tasks to complete autonomy. This can also be seen as a scale from dependence (like for a newborn baby) to independence (of an adult). For machines it looks like a scale from dependence on to the independence from its system design, given knowledge, and a priori knowledge (see Figure 3 ).

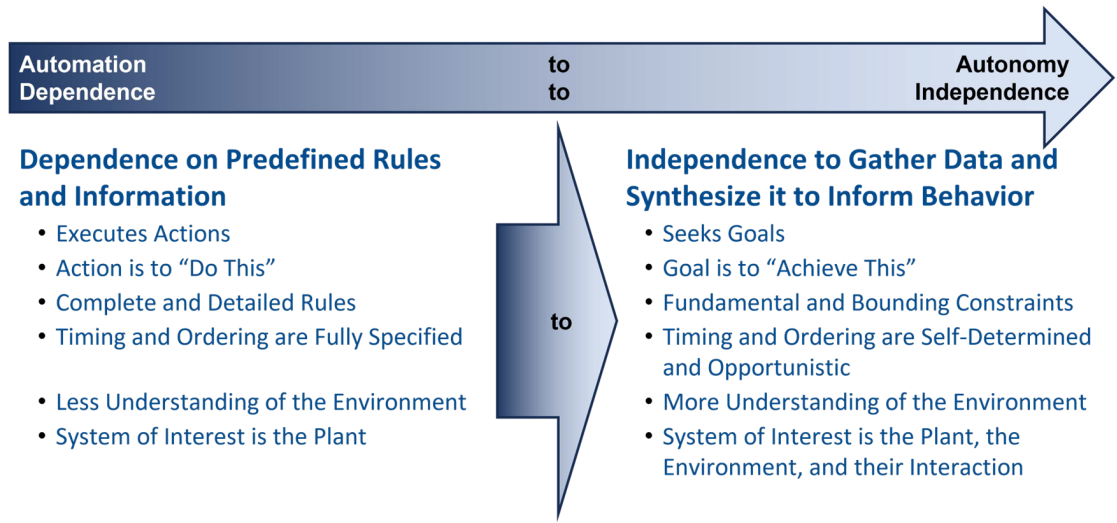

Figure 3. Autonomy on a Sliding Scale

Factory robots replace (some of) the humans doing assembly tasks. Household robots do tasks or chores for (replacing) the humans. The autonomy aspect of a robotic vacuum cleaner is not to vacuum, the machine already does that. The autonomy aspect of a robotic vacuum cleaner is to replace the human's direction about when and where to vacuum.

**The SSCCs**

Along similar thoughts, four primary (high-level) functions (or purposes) of all autonomy systems have been identified: Survival, Success, Collective, and Contextualization of Situations (the SSCCs). Next, these four purposes are described.

**Survival** refers to the fact that all autonomy systems want to act in a way that ensures its own continuation. Autonomy cannot achieve its goals if it ceases to operate. Survival includes all of our normal ideas of safety, health, and well-being as they apply to both humans and machines. In military situations, survival can also imply undetectability and stealth.

**Success** refers to the function of the autonomy system to meet or attempt to meet its mission, goals, and objectives. Its aim to achieve this success is in spite of knowing the system constraints, limitations and potential impacts of decisions. Success gets to the core of the idea of replacing the human by accomplishing the tasks that humans would do.

**Collective** is the third function of an autonomy system. No autonomy agent (human or machine) operates in complete isolation from other agents. Autonomy must work with other elements and agents, and this implies communication among the agents and elements. This working together can serve several purposes: reporting, commanding, and sharing information and/or tasking.

**Contextualizing Situations** is maybe the most difficult function of the autonomy system. Here, the autonomous system is trying to replace the human capabilities of taking advantages of new opportunities, anomaly and hazard detection, extrapolation of past experiences to new situations, reasoning and understanding, problem solving, and developing plans for action, to name a few.

**Working Together**

To be clear, we are promoting a distributed and hierarchical autonomy system. This is in contrast to monolithic (big brain) designs. One aspect of distributed autonomy systems is that they have to work together in order to achieve success. Looking at the kinds of relationships that can exist between autonomy elements (left side of Figure 4), there are fundamentally two types: superior-subordinate and peer-to-peer (right side of Figure 4).

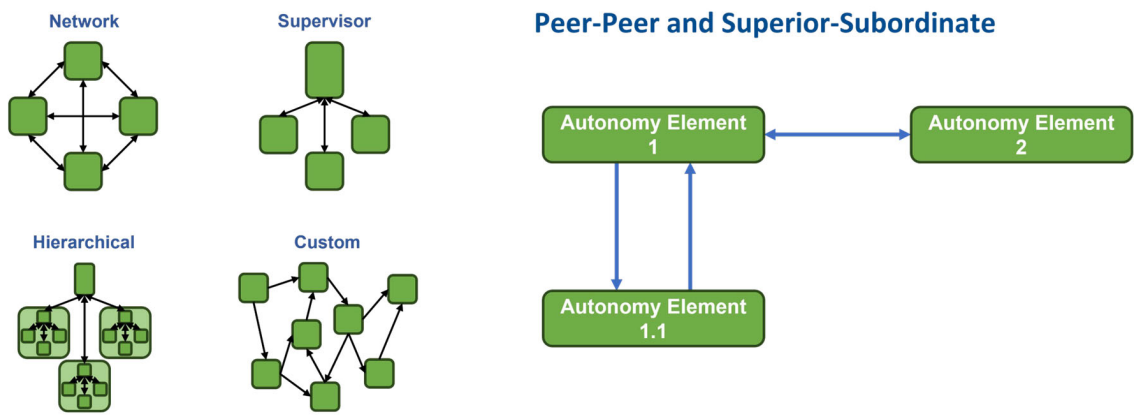

Adapted from: https://langchain-ai.github.io/langgraph/concepts/multi_agent/?ref=blog.premai.io

**Figure 4. Types of Relationships**

Superior-subordinate examples could be that of a manager-employee or parent-child relationships. These examples both typically change over time, moving towards a peer-to-peer type relationship. Peer-to-peer relationships are never truly equal and have some hierarchy in them. So, in the end, both superior-subordinate and peer-to-peer are two extremes of a spectrum and probably no real relationship is truly at either end. This paper focuses on these two extremes to simplify the discussion.

Another important idea is to look at how these elements can interact with each other given these relationships. Next, the four Cs, the four types of interaction, are defined: Cooperation, Coordination, Collaboration, and Command.

**Cooperation** is the lowest level of working together. It includes little beyond sharing information with another element. It does not even care if the receiving element does anything with the information. In cooperation the elements agree on common rules for how to share but there exists no common goal among the elements. Each element has its individual plans and individual execution. A human work-related example would be the simple sharing of calendars. This is the basis behind the publication-subscription (pub-sub) messaging model in networks. The publisher decides what it wants to publish and has no knowledge of any subscribers or their use (or lack of use) of the information.

**Coordination** is the first step up from cooperation. Coordination is focused on either deconflicting or enhancing intersections, as in two people working together to schedule separate meetings so that both can attend. It involves common tasks which is why intersections may occur but no common goal (your meeting could be completely different from my meeting). Coordination can be characterized as joint plans with individual execution. A human example can be seen when multiple professional conferences decide to share a conference location and time to enhance attendance, but each conference is solely responsible for its own sessions and papers. In computers, resource scheduling for shared resources (processor, memory, etc.) is a type of deconflicting for different applications.

**Collaboration** is the next level of engagement. Collaboration involves joint development with common goal(s) and connected tasks. Teams are a good human example. Each person or player may have different roles, but the goal is the same and individual tasks are related. Collaboration can be described as joint plans and joint execution. Distributed algorithms are good examples in machine collaboration.

**Command** is the fourth type of interaction. Command is clearest in the superior-subordinate relationship, where the superior has the authority, responsibility, or role of directing the subordinate. This is considered as a one-way type of interaction and if the subordinate is to report back, they do so as a cooperation interaction. The command interaction allows the superior to override the goal(s) or task(s) of the subordinate. Command can be characterized as individual plan (the superior) and other execution (the subordinate).

Figure 5 shows how the four Cs map to the two types of relationships discussed above.

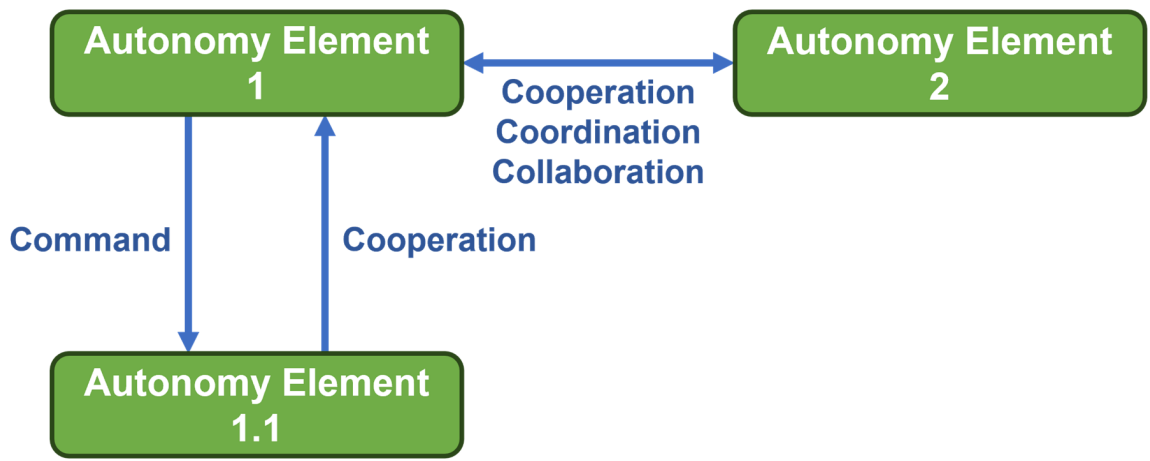

**Figure 5. The Interactions of Relationships**

## LOOPS, LOOPS, AND MORE LOOPS

In [1], Antsaklis and Rahnama make the argument that "Autonomous systems evolve from control systems ...". Specifically, they say:

> "... the system under consideration always has a set of goals to be achieved and a control mechanism to achieve them. This implies that every autonomous system is a control system. Here the term "control system" is used in a most general sense than in the systems and control research area. Control, a decision mechanism that typically uses sensor measurements and feedback together with ways to implement these decisions via actuators, is used to make the system (a very general collection of processes) attain desirable goals."

Using Antsaklis and Rahnama's definition of the term "control system" in the most general sense, one can see that "control loops" exist everywhere. Almost every machine on any kind has control loops embedded in them. This is easy to recognize in vehicle adaptive cruise control systems and thermostats for temperature regulation in buildings, vehicles, and washers and dryers. Control systems are also an embedded part of human systems. Figure 6 contains just a few of the common control loops described in the management and organizational literature. Even the fault detection, isolation, and recovery (FDIR) process can be shown as a loop.

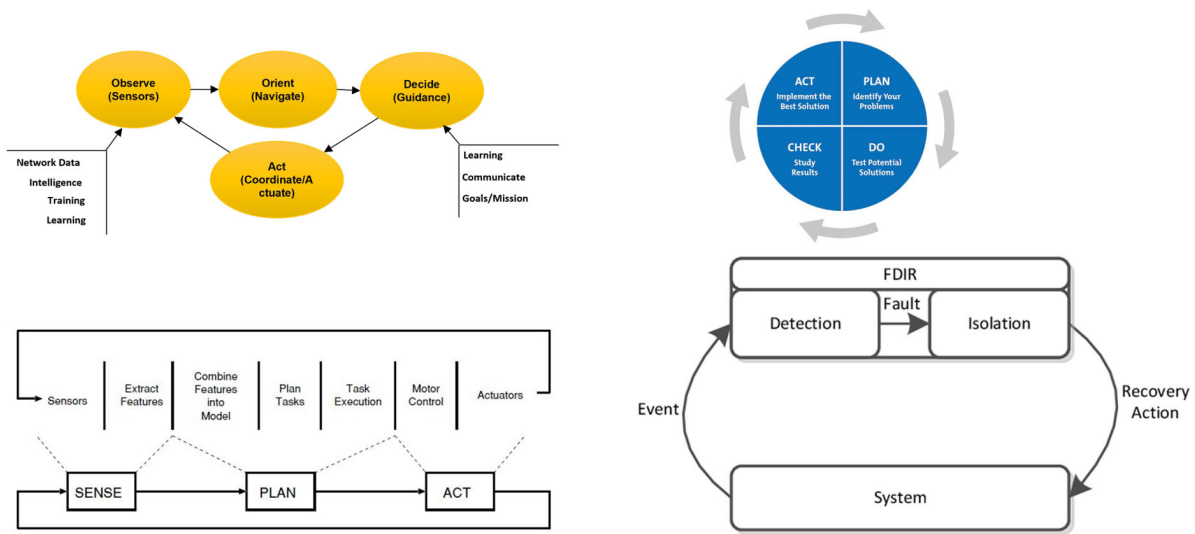

**Figure 6. Example Human System Control Loops ([20], [17], [16])**

Managers at work and parents at home require status updates from employees and children as a way of getting feedback, monitoring progress, and redirecting as necessary. Customer service surveys provide "measurements" of service or product satisfaction. Likes and comments on social media provide valuable insight to content providers, helping to inform their decisions about future content. Basically, the world runs on feedback loops for both machines and humans.

**Control Loop Basics**

Control loops also exist everywhere within a spacecraft. A basic machine-based control loop is shown as a block diagram in Figure 7.

The boxes (capital letters) represent systems or functions. C represents the controller implementation and P is the plant. The word plant is the traditional term used for the system, device, or machine that the controller is controlling. $S_p$ represents the sensor(s) that measure the plant output and $F_p$ the filtering (or fusion or estimation, etc.) applied to the sensor(s) outputs. The filter block's ($F_p$) function is to minimize the effect of the sensor's noise and other measurement inaccuracies and if needed, to fuse or blend to signals of multiple sensors.

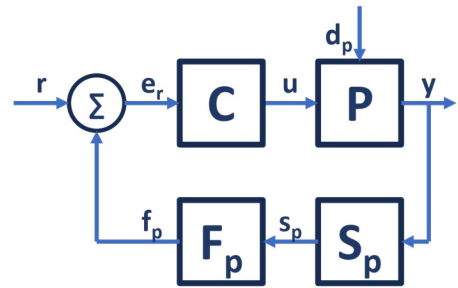

**Figure 7. Basic Control Loop**

The signals are identified in lower case letters and flow from the output blocks to the input of other blocks. The reference input (r) is the desired plant output and $e_r$ is the reference input error. This error is defined as the reference input (r) minus the feedback back signal ($f_p$). the feedback back signal ($f_p$) represents the best estimate of the true plant output or response (y). The plant inputs (u) are the commands the controller uses to drive the plant and $d_p$ is the plant disturbance(s) that also influence the plant output. The measurement of the plant output by the plant sensor(s) is contained in $s_p$. This sensor(s) output also contains sensor noise.

A simple example is the solar array drive subassembly used on spacecraft with independently articulated arrays of panels. The solar panels are used to charge the spacecraft batteries and are ideally pointed at the sun to maximize the electrical current produced by the panels. The control loop is designed so that if the current sensor ($s_p$) sees a decline in current from the panels, the controller (C) will change the array pointing direction by changing the drive signal (u) to the pointing motor (P). This automatic process happens without any continual interaction from outside the loop.

Control loops are typically designed to provide:
- Better performance than the plant alone,
- Stable operation even in the presence of changes to the plant, and
- Robustness to outside disturbances.

Performance considerations are focused on minimizing the magnitude of the reference input error ($e_r$) given a reference input (r). This is called the error rejection capabilities of the loop. Robustness considerations are also focused on minimizing the magnitude of the reference input error ($e_r$) but this time given the plant disturbances ($d_p$). This is called disturbance rejection and is related to the error rejection. The speed of the loop (faster is better) is usually defined as the loop's bandwidth (higher is better) and, among other design options, effects both the error and disturbance rejection capabilities.

Stability considerations are also focused on the magnitude of the reference input error ($e_r$). In this case, for a stable system, an error that converges towards zero as the input changes and that is always finitely bounded is desired. The error would go to zero over time but imperfections in the system, noise, and disturbances limit perfect convergence to zero. The best situation is for the error to remain bounded (stable) by some minimum amount around zero. The smaller the bound the better the performance. The control system is designed to maintain this stability even for (small) changes internal to the plant. As components within the plant change due to temperature and age, their performance characteristic's change. Stability margins are measures of the amount of plant changes that can be accommodated while keeping the whole control loop stable.

**The Outer Autonomy Loop**

The best way to achieve a multi-layered, multi-element autonomy system is to create an outer autonomy loop around every control loop within the spacecraft. As described above, the control loop is focused on controlling the plant, device, or system, with its sensors, performance, stability, and robustness about the plant. An autonomy loop can be wrapped around the control loop as a way to regulate the control loop. In this nested loop configuration (shown in Figure 8 ), the control loop is the inner loop (inside green box), and the autonomy loop is the outer loop.

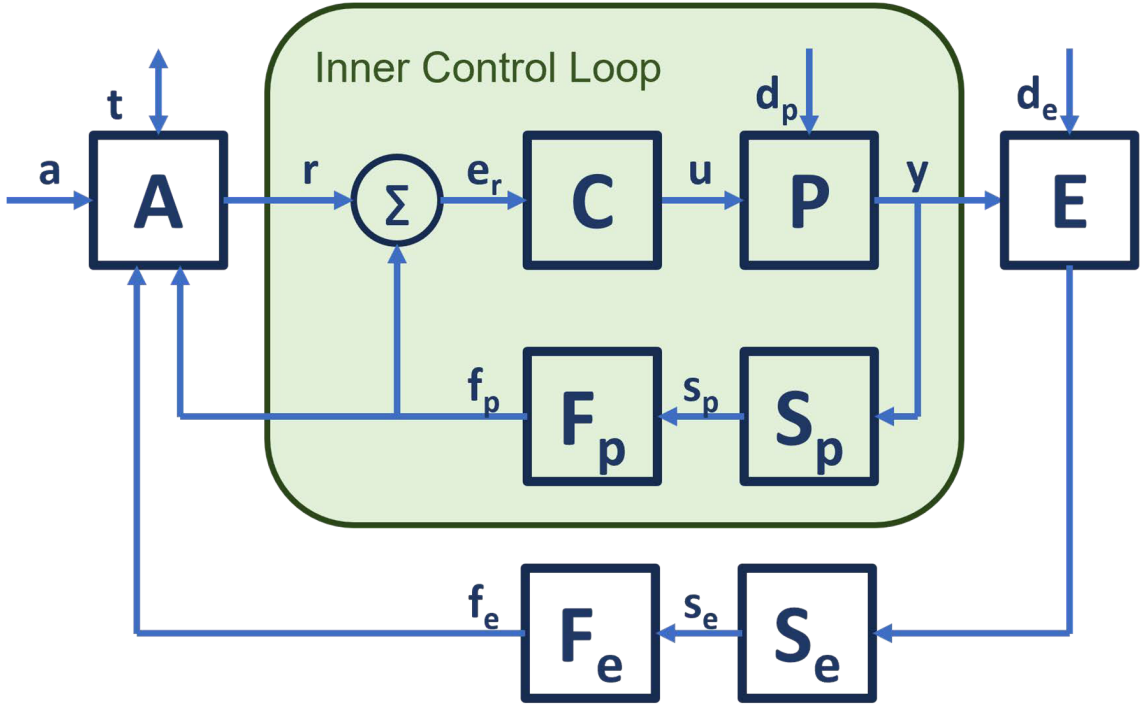

**Figure 8. The Autonomy Loop with Inner Control Loop**

The autonomy loop introduces several new blocks and signals. The block A represents the autonomy function and E represents the environment that the plant is operating in and interacting with. So, from a control's perspective, autonomy (A) is regulating the control loop's action with and in the environment (E). $S_e$ represents the sensor(s) that measure the environment and the plant's interaction in the environment and $F_e$ is the filtering (or fusion or estimation, etc.) of the environment sensor(s) outputs.

The autonomy input signal (a) is the desired plant's action in the environment and may be expressed as a goal or objective. The autonomy block (A) uses the autonomy input (a), the environment feedback ($f_e$), and possibly any inputs from the collective crosstalk channel (t) to determine appropriate reference inputs (r) to the control loop. The autonomy input signal (a) is a command channel as described above in the subsection on the Four Cs of working together. The collective crosstalk channel is used for message passing of the other three C types. It may also be desirable for autonomy (A) to have access to the plant feedback ($f_p$) and so this is also shown. The environment may also have disturbance(s) acting on it ($d_e$). The measurement of the environment by the environment sensor(s) is contained in $s_e$. This sensor(s) output also contains sensor noise.

The outer autonomy loop is then different from the control loop because the autonomy loop is focused on the environment and the plant's interaction in and with the environment. The control loop is focused solely on the plant. The autonomy loop's sensors, performance, stability, and robustness are about the environment and interaction in and with environment.

For a simple autonomy loop example, let's consider the same solar array drive subassembly as before. One aspect of the autonomy block could be to receive messages (through the collective crosstalk channel (t)) to inform it of times when the satellite is going into Earth's shadow and will be blocked from receiving sunlight. During these times the control loop would have an impossible task of finding the right pointing direction to maximize generated current. The autonomy block could simply command the controller into a cage mode. The cage mode locks the array pointing to a specific direction without regard to the generated current. The autonomy block simply commands the controller back to normal operations once the satellite passes back out of Earth's shadow.

The difference between control loops and autonomy loops can be summarized in the following table (Table 1).

**Table 1. Control and Autonomy Loop Differences**

| Control Loops | Autonomy Loops |
|---|---|
| Control the Plant (a Part of the System) | Control the System's Interactions in the Environment |
| Sensors Measure Plant Motion | Sensors Monitor Activity in the Environment |
| Actuators Move the Plant | Actuators Orchestrate Actions of the System in the Environment |
| Performance is Minimizing Errors in Plant Motion | Performance is Minimizing Deviations for the Plan/Goal |
| Stability is Maintaining Bounded Plant Motion Errors | Stability is Maintaining Convergence Towards the Plan/Goal |
| Robustness is Minimizing Errors due to External Disturbances | Robustness is Minimizing Errors due to Unknown Situations (Surprise) |

**EXAMPLES**

Before proceeding, a space system taxonomy, is introduced, to be able to better discuss some examples of Autonomy at Levels within a spacecraft.

**A Space System Taxonomy**

As already mentioned, we advocate embed autonomy loops at all levels within a spacecraft decomposition. In Figure 9, an outline for this spacecraft decomposition is shown.

The decomposition dimension is typical of many space systems. The Factor dimension may be new to some, and Figure 9 only has two of the factor columns filled out: the WBS/PBS and Functional columns. Figure 9 also contains a brief description of each level in the decomposition with some examples. The Functional factor column not only labels the kinds of functional descriptions that go into each cell of the column but also links the appropriate "measures" to each level. Measures of Objective (MOO) are at the Mission level, Measures of Effectiveness (MOE) are at the Element level, Measures of Capability (MOC) are at the Segment level, Measures of Performance (MOP) are at the System level, and Technical Performance Measures (TPM) are at the Subsystem level. The Interfaces factor is meant to identify the connections a particular element has with other elements. The Configurational factor is to identify all configuration parameters of the element in question. Finally, the Procedural factor is to identify the appropriate ordering of steps to taken within the element.

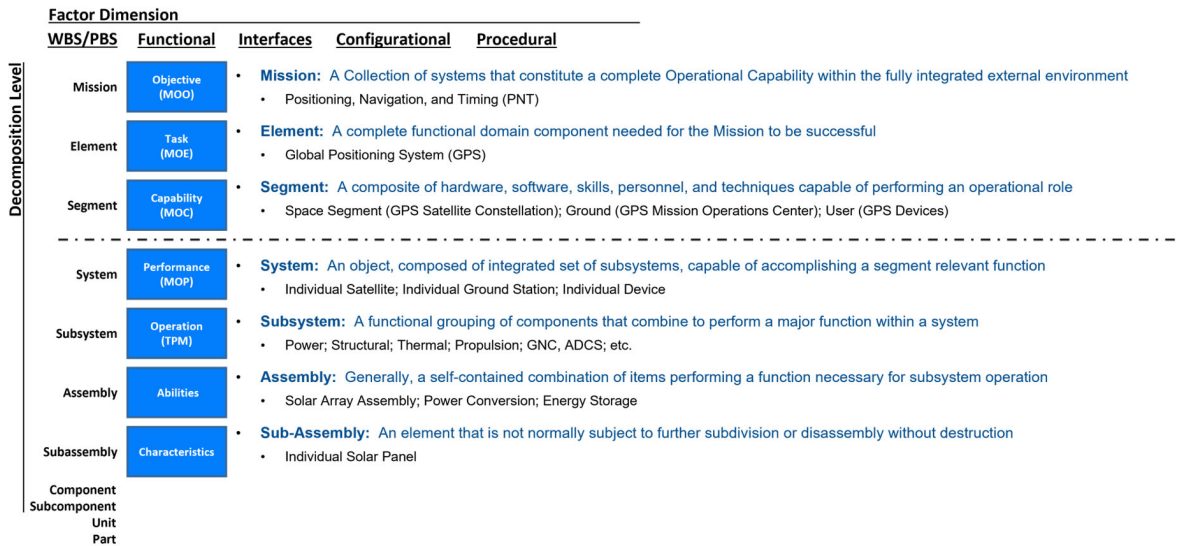

Figure 9.  A Space System Taxonomy

The following three examples are used to illustrate how Autonomy at Levels could be implemented at three different levels within a space system taxonomy. For each example, the control loop's environment and the SSCCs, the four primary functions of autonomy are discussed, as a way of determining each autonomy loop's specific tasks. The examples come from a simple power subsystem decomposition as summarized in Figure 10.

| | WBS/PBS | Functional | Interfaces | Configurational | Procedural |
|---|---|---|---|---|---|
| Mission<br>• Objective<br>• MOO | | | | | |
| Element<br>• Task<br>• MOE | | | | | |
| Segment<br>• Capability<br>• MOC | Satellite Cluster or Constellation | | | | |
| System<br>• Performance<br>• MOP | Satellite | | | | |
| **Subsystem**<br>• **Operation**<br>• **TPM** | Power | Provide Electrical Energy to Components<br>• Generation<br>• Storage<br>• Distribution | • Satellite<br>• Payload<br>• … | • Normal<br>• Energy Saving<br>• … | |
| **Assembly**<br>• **Abilities** | Generation | Solar Panels to Collect and Generate<br>• Solar Panels<br>• Solar Drive | • Power<br>• Storage, Dist.<br>• Panels, Drive | • Charging<br>• Off<br>• … | |
| **Subassembly**<br>• **Characteristics** | Solar Drive | Point Panels at Sun to Maximize Generation | • Motor<br>• Sensor | • Normal<br>• Stowed | |
| **Component** | Circuit Breaker | Protect Drive from Drawing Excess Current | • Switch | • Closed/Open | |

Figure 10.  Power Subsystem Examples

The top four levels of the space system taxonomy are grayed out because they will not play a role in our examples. The procedural column is also left blank as these are meant to contain detailed ordering of steps to be taken for different configurations. These details are beyond the scope of this paper and so are omitted.

The power subsystem is only one of many subsystems within a satellite. In these simplified examples, the power subsystem is broken down into three assemblies: generation, storage, and distribution. For the examples here, generation can be broken down into the solar array subassembly and the solar array drive subassembly. The final level breakdown is the solar array drive subassembly into its motor, sensor, and circuit breaker components.

The examples start from the bottom and work up. Each example will be very idealized and simplified in an attempt to highlight the basic Autonomy at Levels concepts without getting too bogged down in technical design details.

**Example 1 - Circuit Breaker**

The first simple example is that of a circuit breaker. In terrestrial electrical systems, it is common to have a fuse or circuit breaker embedded within important or critical circuits. In spacecraft, the circuit breaker could be implemented as a commandable switch within the circuit. The switch would be commandable so that if tripped, it can be reset via remote commands. Referring back to Figure 7, the control loop for the circuit breaker can be described with the switch (opening and closing mechanism) as the plant, a position sensor that reads the switch as either in the open or closed position (no filtering needed for this example), and the controller that commands the switch position and confirms the correct position when the reference input error is zero. The reference input is the desired open or close position.

A commanded open position would be the normal operating condition. In situations where the electrical component(s) in the circuit are to be shutdown, the switch position is closed. Referring back to Figure 8, to add an autonomy loop around this example control loop, the first step would be to define the environment. At very low-levels within the power subsystem breakdown, the control loop is very simple and the environment it operates in is also simple. In this example, the environment can be defined as the current in the circuit. This means that the environmental sensor of interest is a current sensor to measure the flow of electricity in the circuit. The filtering block could be a simple low-pass filter to smooth out the high frequency sensor noise. With the measured current as the autonomy feedback signal, the autonomy block can be used to automatically send an open command to the controller if the circuit's current exceeds some threshold value. This is the Survival function of the autonomy loop described in the SSCCs above and this now constitutes an automatic circuit breaker.

At this point, it may be obvious to state, that most spacecraft already have this functionality built into their power subsystems. Good, they are already applying Autonomy at Levels.

When considering the Collective function of autonomy, the autonomy block can subsume the role of communicating status and faults in the circuit up to higher-level autonomy elements and this is done through the collective crosstalk channel. Since this is an example at a low-level within the power subsystem breakdown, the situations that need contextualizing are also very simple. The "understanding" that the autonomy block needs is about present electrical current levels and maximum permissible levels. This "understanding" is simple enough to be encoded in a state machine with transition criteria and resulting state actions.

Next, consider that the circuit being discussed is that which charges spacecraft batteries, then one would want to add the charge state of the batteries to the environment. Now, it is illustrative to talk about the Success function of the SSCCs. In this case, success can be defined as the normal operation of closing the switch when the batteries need charging and opening the switch when they are fully charged. A battery charge state sensor needs to be added to the environment feedback path and more state machine logic added to the autonomy block to handle the additional state, transitions, and actions.

One last thing to discuss, the autonomy input signal (a). This is the command to the autonomy block and it takes on a different meaning than the reference input (r) to the control loop. The autonomy input would represent one of the different possible operating modes of the autonomy loop. In addition to the normal operating modes, discussed so far, an emergency safe mode could be defined where the autonomy block shuts down nonessential components. Another possible aspect to this command could be to adjust parameters with the autonomy's state machine logic. An example where this might be needed, is the need to increase the current threshold for an aging reaction wheel. Still another possible aspect to this input could be the ability of a higher-level autonomy element to temporarily override the basic functioning of this loop. This kind of example would be analogous to our brain overriding out reflex reaction to let go of a hot object if letting go were to let the object hit another person.

These lower-level autonomy loops can easily be thought of as the spacecraft's reflexes and their inner control loops as the regular, normal operations that happen in the background as automatic functions without the need for continuous attention from higher-level autonomy elements.

**Example 2 - Solar Array Drive Subassembly**

In the previous section of this paper, the solar array drive subassembly example was started. This discussion is now added to and completed. To recap a bit, the solar panels are used to charge the spacecraft batteries and are ideally pointed at the sun to maximize the electrical current produced by the panels. The control loop is designed so that the array of panels pointing direction is constantly adjust to maximum electrical current production. The plant, for this system, is the drive motor together with its solar array.

The primary element in the environmental that the autonomy loop should be interested in is the Sun, since that is the source of our example's electrical power. As alluded to before, there may be times when sunlight will not be available. The control loop is totally unaware of this as its sensors are focused on the current the solar panels are producing. The autonomy loop can account for this by either having a sun sensor in its feedback path or messages coming in from some other spacecraft element that is tracking the position of both the Sun and the Earth and able to predict when the satellite will be in shadow. As a side note, other subsystems, like thermal, will also be interested in knowing the times for illumination and shadow. These messages can be thought of as virtual sensors to help in continuing to use the autonomy loop paradigm.

There are other possible environmental elements that may be important for specific spacecraft designs and missions. For example, solar panel temperatures or oscillations and vibrations induced by the motor on the spacecraft when driving the arrays. Sensors would need to be incorporated into the feedback path as input into the autonomy block. The "understanding" and action logic to address these possibilities would need to be incorporated into the autonomy block's functionality.

The Survival function of the autonomy does not need to worry about over current situations burn-ing out the sensor or motor as the lower-level circuit breaker autonomy loop has that handled. One possible survival concern, at this level within the power subsystem decomposition, could be solar flares and particle storms that could damage the solar panels. If some kind of sensor or information source is available for these threats, then the autonomy loop could react. One possible reaction could be to command the controller to safe or stowed configuration, until normal operations were safe to continue.

As already highlighted, there is more Collective cooperation than at lower-levels. The Collective function could also be producing outgoing messages with status and faults as well as operating modes changes (like cage mode) that could impact other spacecraft elements. The Contextualizing Situations function is a little more complicated as a broader variety of situations can be encountered and understanding the mix of inputs a bit more challenging. The Success function would involve getting all of the commands and messages needed and responding with appropriate actions. At this level, most of the autonomy functions can still be handled with decision tress and state machine logic. It is also recognized that much of the functionality discussed in this example is also already being implemented on spacecraft. Good news, Autonomy at Levels is already working.

The autonomy input signal would represent one of the different possible operating modes of the autonomy loop. The input could also be used for a higher-level autonomy element to temporarily override the basic functioning of this loop. A situation where this might be relevant is when the satellite's payload is in the middle of performing its primary mission and does not want to be interrupted by the solar drive array transitioning to cage mode.

**Example 3 - Power Generation**

The power generation assembly may not have a traditional control system, but still has a control loop as generalize in the previous section. The generation control loop functions will probably be implemented in a state machine which is getting charging feedback from the storage assembly and component use feedback from the distribution assembly. Inputs are probably mode switching commands.

Autonomy loop inputs from the power subsystem will probably look like schedules for spacecraft mode changes to support mission timelines and maybe usage demands forecasts. Usage demands forecasts are a possible type of higher-level "thinking" (analysis) that might happen at the power subsystem level.

Part of the Success and Survival autonomy functions are the health and well being of the components within the generation assembly: the solar arrays, drive motors, current sensors, etc. Lower-level autonomy loops have handled some of the fault conditions and threats but from the higher-level generation assembly perspective a more holistic picture of the health of generation can be viewed. This could include functions to look at slow component degradation (e.g. solar cells) and possible mitigation or preventive maintenance type actions. As alluded to before, possible model parameter changes could be detected ("learned") and lower-level components using those models could then be commanded to update those parameters.

If characteristics are degrading or changing over time, then the Collective function would be used to communicate these changes to other potentially affected autonomy elements. The satellite system level could use this information to update mission operational steps and actions to accommodate the changes.

It may also be clear that at this power generation assembly level the Contextualizing Situations functions of the SSCCs is more complicated. The autonomy loop must juggle energy and power consumption needs with its generation capabilities while factoring in health and status and threats and faults. Thus, contextualizing situations is more complicated because the environment for generation is a large field of concern. The situation for satellite system level is even bigger as the satellite will be concerned about both the entire internal environment of the spacecraft as well as the external orbit environment.

It is also possible, back at the power generation assembly level, that this maybe where some of the temporary override commands come from. Seeing some individual component's temperature rising, it could "choose" to temporarily overload another circuit to provide some relief. In a hierarchical autonomy system like this, a higher-level element could even allow or mandate the demise (sacrifice) of a lower-level element for survival or success of the larger system. An example of the sacrifice of a lower-level element could be to allow a section of solar cells to burn out during a unusual but particularly demanding and critical mission.

**SUMMARY**

This paper advocates the use of Autonomy at Levels concepts for the development of autonomy functions within a spacecraft. We have tried to explain the basic ideas and concepts involved and given some examples at different levels within a spacecraft's system decomposition. These examples are not complete and are meant to be illustrative and not prescriptive. The key ideas include:
- Push autonomy functions down to the lowest levels possible,
    - This allows for more compute time at the higher-levels for "thinking"
- Look to create an autonomy loop around every control loop,
    - Let the control loop handle normal operations
- Define the environment that each autonomy loop is concerned with,
    - And the measurement feedback from the environment
- Use the SSCCs for each autonomy loop to ensure complete autonomy coverage for the given level and element,
    - Survival
    - Success
    - Collective
    - Conceptualizing Situations
- Define the autonomy inputs and their formats, and
- Define the connections and messages being passed on the collective crosstalk channel.

**REMAINING QUESTIONS AND FUTURE RESEARCH**

This paper is only an introduction to the Autonomy at Levels paradigm. Many questions have not been answered here and are aspects for future work. A few of these questions are listed here.
- In using the Systems Engineering V, what do the top-level autonomy requirements look like?
- What important ideas should be included in these top-level requirements?
- How to identify and describe these autonomy requirements for a spacecraft or a constellation?
- How to use the autonomy requirements to identify and describe what autonomy functions need to exist within a system or spacecraft?
- How do those requirements get flowed down to system and subsystem level specifications so that detailed embedded autonomy designs can follow?
- Can autonomy functions be defined that are applicable at each level in a system taxonomy?

• Are these functions, roles, capabilities, or skills or some combination?

• How to design autonomy elements within a system or spacecraft, if a requirement were to exist?

• Can a Generic Autonomy Template be developed to be applied to any level to assist in the requirements, flow down and design process in the system taxonomy?

• During integration and testing, how are autonomy functionality and performance evaluated?

• How do the autonomy elements roll-up to form higher level capabilities and how are those tested and evaluated?

• Are other verification methods needed in addition to the standard: inspection, demonstration, analysis, and test?

• Maybe there is the need for independent simulations or some kind of comparison to human performance?